%% file: main.tex
\documentclass[lettersize,journal]{IEEEtran}
\usepackage{amsmath,amsfonts}
\usepackage{algorithmic}
\usepackage{array}
\usepackage[caption=false,font=normalsize,labelfont=sf,textfont=sf]{subfig}
\usepackage{textcomp}
\usepackage{stfloats}
\usepackage{url}
\usepackage{verbatim}
\usepackage{graphicx}
\usepackage[numbers]{natbib}
\usepackage{multirow}
\def\BibTeX{{\rm B\kern-.05em{\sc i\kern-.025em b}\kern-.08em
    T\kern-.1667em\lower.7ex\hbox{E}\kern-.125emX}}
\usepackage{balance}
\usepackage[most]{tcolorbox}
\usepackage{xspace}
\usepackage{tabularx}

\newcommand{\blue}[1]{\textcolor[rgb]{0.00,0.00,1.00}{#1}}
\newcommand{\red}[1]{\textcolor[rgb]{1.00,0.00,0.00}{#1}}

\newcommand{\DA}[0]{V2P dataset\xspace}
\newcommand{\DB}[0]{V2N dataset\xspace}
\newcommand{\DAShort}[0]{V2P\xspace}
\newcommand{\DBShort}[0]{V2N\xspace}
\newcommand{\framework}[0]{TrapEval\xspace}

\newcounter{findingCounter}

\newcommand{\finding}[2]{
  \begin{tcolorbox}[enhanced, left=3mm,right=3mm,
    colback=gray!10, colframe=gray!80, boxrule=0pt,
    borderline west={4pt}{0pt}{gray!90},
    breakable
    ]
    \stepcounter{findingCounter}
    \textbf{Finding \thefindingCounter:} #2
    \end{tcolorbox}
}

\begin{document}
\title{Do Fine-Tuned LLMs Understand Vulnerabilities? An Investigation into the Semantic Trap}
\author{Feiyang~Huang,
        Yuqiang~Sun,
        Fan~Zhang,
        Ziqi~Yang,
        Han~Liu,
        and~Yang~Liu%
\thanks{F.~Huang, F.~Zhang, and Z.~Yang are with the College of Computer Science and Technology, Zhejiang University, Hangzhou, China (e-mail: feiyanghuang@zju.edu.cn; fanzhang@zju.edu.cn; yangziqi@zju.edu.cn).}%
\thanks{Y.~Sun and Y.~Liu are with Nanyang Technological University, Singapore (e-mail: yuqiang.sun@ntu.edu.sg; yangliu@ntu.edu.sg).}%
\thanks{H.~Liu is with the College of Cryptology and Cyber Science, Nankai University, Tianjin, China (e-mail: hanliu@nankai.edu.cn).}%
\thanks{Corresponding author: Fan Zhang.}}

\maketitle

\begin{abstract}

Large Language Models (LLMs) have shown promising performance in software vulnerability detection, particularly after domain-specific Supervised Fine-Tuning (SFT). However, it remains unclear whether these models genuinely internalize vulnerability root causes or merely exploit surface-level functional patterns shared across similar code. While prior work has documented related failures on pre-trained encoder models or zero-shot prompting of decoder-only models, the SFT process itself, and how explicit reasoning supervision modulates it, has not been systematically dissected. In this paper, we focus on this gap: we study fine-tuned decoder-only LLMs under both vanilla SFT and SFT with distilled reasoning supervision, and we identify a failure mode that we term the \textit{Semantic Trap}, characterized by three observable symptoms: pairing-sensitive performance, gap-dictated decisions, and fragility to semantic-preserving change.
To probe these symptoms, we propose TrapEval, an evaluation framework comprising two datasets derived from real-world projects, V2P (pairing vulnerable code with its patched version) and V2N (pairing vulnerable code with unrelated normal code), together with cross-dataset testing, semantic-preserving perturbations, CodeBLEU-based semantic-gap analysis, and an LLM-assisted taxonomy of reasoning failures. Using TrapEval, we fine-tune five representative state-of-the-art LLMs under two paradigms: with and without explicit reasoning (i.e., Chain-of-Thought).
Our empirical results show that fine-tuning without explicit reasoning produces deceptively high scores on unpaired data (V2N) while failing under all three symptoms: models suffer large false-positive rates on patched code (V2P), degrade under semantic-preserving perturbations, and exhibit a weak but statistically systematic dependency on the textual gap between vulnerable and patched code (Spearman $|\rho|<0.1$, $p<0.05$ across all five models). Fine-tuning with explicit reasoning reduces these symptoms but at the cost of recall, and its absence of measurable gap-dependency is partly compatible with a floor effect rather than direct evidence of escaping the trap. Furthermore, our reasoning-failure taxonomy reveals that explicit-reasoning models still frequently misinterpret control flow and hallucinate API behavior, indicating that current fine-tuning practices mitigate but do not eliminate reliance on surface-level features.
\end{abstract}

\begin{IEEEkeywords}
Large Language model, Fine-Tuning, Robustness, Vulnerability Detection.
\end{IEEEkeywords}

\input{Sections/1-Intro}
\input{Sections/2-Background}
\input{Sections/3-Study}
\input{Sections/4-Experiments}
\input{Sections/5-Discussion}

\input{Sections/7-Conclusion}

\balance

\bibliographystyle{IEEEtran}
\bibliography{sample-base}

\end{document}

%% file: Sections/1-Intro.tex
\section{Introduction}

The rapid evolution of Large Language Models (LLMs) has fundamentally reshaped the landscape of software security, particularly in critical tasks such as vulnerability detection, localization, and automated repair~\cite{sheng2025llms, zhou2025large, shi2025cgifuzz}. To fully harness the potential of these models, researchers have extensively explored various LLM-based techniques specifically tailored for vulnerability analysis.
Efforts~\cite{sun2024gptscan, david2023you, cao2024llm, wen2024scale, lu2024grace, sindhwad2025vulnarmor} have yielded impressive performance, often surpassing static and dynamic analysis tools. Beyond utilizing off-the-shelf foundation LLMs, a significant line of recent work \cite{ma2024combining, du2024generalization, boi2024smart, yang2024large, zhang2024empirical} has shifted towards developing domain-specific vulnerability detection models. By employing Parameter-Efficient Fine-Tuning techniques~\cite{xu2023parameter, li2021prefix, hu2023llm, lester2021power, petrov2023prompting}, most notably Low-Rank Adaption~\cite{hu2022lora}, researchers have successfully adapted general-purpose LLMs to specialized security datasets, further pushing the performance boundaries of automated vulnerability management.

As the research community increasingly adopts LLMs for vulnerability-related tasks, significant concerns regarding their reliability and robustness have surfaced~\cite{ullah2024llms, li2025sv, li2025everything, yin2024multitask, jiang2024investigating}. 
Early extensive evaluations of "out-of-the-box" pre-trained models revealed that while they show promise, they exhibit high sensitivity to minor semantic variations and a lack of determinism~\cite{ullah2024llms}.
More critically, a recent line of research has begun to investigate what these models actually learn, uncovering a pervasive lack of logical robustness~\cite{ding2024primevul, li2025sv, risse2024uncovering, weissberg2025llm}.
For instance, recent benchmarks~\cite{ding2024primevul} utilizing vulnerability-patch pairs demonstrate that LLMs consistently fail to distinguish a vulnerable code snippet from its corresponding patched version. Further probing studies indicate that models often suffer from severe overfitting to label-unrelated features (e.g., variable names, unexecuted statements) and rely heavily on surface-level statistical characteristics rather than inherently mining deep semantic logic~\cite{risse2024uncovering, weissberg2025llm}. Collectively, these findings suggest that off-the-shelf models largely rely on functional pattern matching rather than rigorous security reasoning.

However, all of these probing efforts target either pre-trained encoder models~\cite{risse2024uncovering} or zero-shot prompting of decoder-only LLMs~\cite{ding2024primevul, li2025sv, weissberg2025llm}; none of them dissects the supervised fine-tuning process itself, which is the very mechanism most modern security pipelines rely on to specialize LLMs for vulnerability detection. Whether SFT and in particular SFT with explicit reasoning supervision repairs, leaves intact, or even amplifies this reliance on surface patterns is therefore an open question. Our work is the first targeted study of this question, and it provides the empirical anchor for the failure mode we name the \textit{Semantic Trap} below.

To compensate for these out-of-the-box limitations, the community has widely turned to Supervised Fine-Tuning (SFT) to adapt general-purpose LLMs into domain-specific security experts~\cite{boi2024smart, ma2024combining, du2024generalization}.
While fine-tuning is widely treated as a ``black-box'' performance booster, the current literature fails to scrutinize whether this process helps models truly internalize the \textit{vulnerability root cause} (the intrinsic security flaw) or merely encourages them to exploit \textit{functional patterns} (surface-level features shared across functionally similar code) as a shortcut.
We refer to this failure mode as the \textit{Semantic Trap}, and we characterize it operationally rather than as a vague preference for particular code domains. Concretely, we say a fine-tuned LLM is trapped when its vulnerability detection behavior exhibits the following three observable symptoms:
(i) \textbf{Pairing-sensitive performance.} The model attains high F1 when a vulnerable function is contrasted with a semantically unrelated benign function, yet its performance collapses when the negative sample is the corresponding patched version that differs only in security-critical logic.
(ii) \textbf{Gap-dictated decisions.} On (vulnerable, patched) pairs, the model's detection performance is statistically explained by the textual--syntactic distance between the two snippets (e.g., as measured by CodeBLEU) rather than by the underlying security semantics, so a shrinking gap predictably degrades accuracy.
(iii) \textbf{Fragility to logic-preserving change.} The model's predictions degrade under semantic-preserving perturbations that leave the security logic untouched, indicating reliance on lexical or structural cues co-occurring with, rather than constituting, the vulnerability.
A model exhibiting these symptoms behaves as a sophisticated pattern discriminator: it can flag code that ``looks like'' vulnerable code, but it cannot reliably reason about whether a specific vulnerability is actually present.
Given the risk of deploying detectors whose high benchmark scores may not reflect genuine security understanding, our work aims to open this fine-tuning black box.
We investigate whether contemporary SFT practices, both vanilla SFT and SFT enhanced with explicit reasoning (i.e., Chain-of-Thought, CoT), actually help models comprehend vulnerabilities, or whether they primarily train models to act as efficient pattern discriminators.

To investigate this phenomenon and assess the depth of LLM understanding, we propose TrapEval, an evaluation framework designed to disentangle vulnerability root cause from functional pattern. Our framework begins with the construction of two datasets derived from real-world open-source projects: the V2P dataset (Vulnerable-to-Patch pairs), which forces models to distinguish near-identical code that differs only in subtle security logic, and the V2N dataset (Vulnerable-to-Normal pairs), representing traditional unpaired detection benchmarks. Using TrapEval, we fine-tune five representative state-of-the-art LLMs (including the Qwen~\cite{qwen3technicalreport}, Llama~\cite{grattafiori2024llama}, and DeepSeek~\cite{deepseekai2024deepseekv3technicalreport} families) under the two paradigms defined above (with or without explicit reasoning). We then conduct a multi-dimensional evaluation encompassing cross-dataset testing, semantic-preserving data augmentations, CodeBLEU-based~\cite{ren2020codebleu} semantic gap analysis, and an LLM-assisted taxonomy of explicit-reasoning failures.

To summarize, the primary contributions of this paper are summarized as follows:
\begin{itemize}[]
    \item We constructed two specialized datasets, \DA (Vulnerable-to-Patch) and \DB (Vulnerable-to-Normal), derived from real-world open-source projects. These datasets are specifically designed to decouple security logic from the functional pattern, providing a rigorous benchmark to quantify whether LLMs are learning the intrinsic logics of a vulnerability or are merely being misled by the functional patterns of the code.
    \item  We developed \framework, an evaluation framework that includes dataset construction, parallel fine-tuning processes  (with and without explicit reasoning), and multi-dimensional evaluation to systematically probe whether a fine-tuned LLM for vulnerability detection only focuses on functional patterns.
    \item We conducted an extensive evaluation of five LLMs. Our findings provide empirical evidence that current standard fine-tuning practices often fail to impart vulnerability reasoning, instead encouraging models to rely on functional patterns, a distinction that future LLM-based security research must address. Furthermore, our in-depth analysis reveals that even when models are forced to produce explicit reasoning, they still frequently fail in ways consistent with shallow semantic understanding, such as misinterpreting control flow or hallucinating API behavior.

\end{itemize}

%% file: Sections/2-Background.tex
\section{Background \& Related Works}

\subsection{Related Works}
 
\textbf{LLM for Vulnerability Task. } The rapid evolution of decoder-only LLMs has reshaped vulnerability detection research, with many recent studies adopting LLMs as core analysis engines~\cite{sheng2025llms, zhou2025large}.
To compensate for the limited ability of general-purpose LLMs to reason about program semantics, prior work has explored various augmentation strategies.
A prominent line of research integrates LLMs with Retrieval-Augmented Generation (RAG) to supply external security knowledge or historical vulnerability context~\cite{cao2024llm, keltek2025lsast, kouliaridis2024assessing, mathews2024llbezpeky}.
Other approaches combine LLMs with structural program representations, such as Abstract Syntax Trees~\cite{zhou2024comparison, mao2024towards, wen2024scale, sindhwad2025vulnarmor}, and data-flow and control-flow analyses to expose execution-level vulnerability cues~\cite{wang2023defecthunter, mahyari2024harnessing, li2024iris, liu2024exploration, lu2024grace}.

Beyond external augmentation, Supervised Fine-Tuning has emerged as a key technique for embedding domain-specific security knowledge directly into LLMs.
Prior work shows that fine-tuning can improve detection performance in specialized settings.
For example, Boi et al.~\cite{boi2024smart} fine-tune LLMs for smart contract security using structured vulnerability taxonomies, while iAudit~\cite{ma2024combining} adopts a two-stage fine-tuning framework combined with multi-agent reasoning to enhance audit accuracy and interpretability.
VulLLM~\cite{du2024generalization} further explores multi-task instruction fine-tuning to jointly support vulnerability localization and explanation, improving generalization across datasets.

\textbf{Evaluation of LLMs in Vulnerability Detection.} With the rapid advancement of LLMs, evaluating their effectiveness in code vulnerability detection has attracted significant attention. 
Early studies primarily examined the out-of-the-box capabilities of pre-trained LLMs. 
Large-scale benchmarks by Khare et al.~\cite{khare2025understanding} and Steenhoek et al.~\cite{steenhoek2024err} show that although LLMs demonstrate promising detection performance, they are highly sensitive to minor semantic variations.
Ullah et al.~\cite{ullah2024llms} further report non-determinism and limited reliability in model predictions. 
To improve performance, subsequent work explored enhanced prompting and contextual information~\cite{li2025everything}, while others shifted toward understanding the internal mechanisms of LLMs from a modular perspective~\cite{sun2024llm4vuln, yin2024multitask, jiang2024investigating}.

Recently, a critical line of research has begun to investigate what these models actually learn, revealing a pervasive lack of logical robustness. For instance, recent benchmarks such as PrimeVul~\cite{ding2024primevul} and SV-TrustEval-C~\cite{li2025sv} demonstrate that LLMs consistently fail to distinguish vulnerable code from its corresponding patched version, suggesting a reliance on functional patterns rather than security-critical logic. Delving into the root causes of such failures, Risse et al.~\cite{risse2024uncovering} demonstrated that machine learning models for vulnerability detection suffer from severe overfitting to label-unrelated features (e.g., variable names, unexecuted statements) and lack out-of-distribution (OOD) generalization when subjected to semantic-preserving transformations. Similarly, Weissberg et al.~\cite{weissberg2025llm} evaluated LLMs through the lens of traditional code metrics, concluding that LLMs do not inherently mine deep semantic logic; rather, their predictions heavily rely on surface-level statistical features that are equivalent to simple syntactic code metrics.

\begin{table}[t]
\centering
\caption{Positioning of our study relative to representative prior work on the reliability of LLM-based vulnerability detection. ``$\bullet$''~=~fully addressed; ``$\circ$''~=~partially addressed; ``--''~=~not addressed.}
\label{tab:positioning}
\small
\setlength{\tabcolsep}{3.5pt}
\renewcommand{\arraystretch}{1.15}
\begin{tabular}{l|cccccc}
\hline
\textbf{Study} & \textbf{D-LLM} & \textbf{SFT} & \textbf{V2P} & \textbf{Gap}  & \textbf{Tax.} \\
\hline
PrimeVul~\cite{ding2024primevul}         & $\bullet$ & --        & $\bullet$ & --        & -- \\
SV-TrustEval-C~\cite{li2025sv}           & $\bullet$ & --        & $\bullet$ & --      & -- \\
Risse et al.~\cite{risse2024uncovering}  & --        & $\circ$ & $\bullet$        & --      & -- \\
Weissberg et al.~\cite{weissberg2025llm} & $\bullet$ & --        & --        & $\circ$    & -- \\
\hline
\textbf{Ours (\framework)}               & $\bullet$ & $\bullet$ & $\bullet$ & $\bullet$  & $\bullet$ \\
\hline
\end{tabular}

\vspace{3pt}
\raggedright\footnotesize
\textbf{D-LLM}: coverage of contemporary decoder-only LLMs (as opposed to encoder-based models such as CodeBERT/UniXcoder).\;
\textbf{SFT}: systematic analysis of the supervised fine-tuning process (with/without explicit reasoning), rather than treating fine-tuning as a black-box performance booster.\;
\textbf{V2P}: paired vulnerable--patch evaluation that forces models to discriminate near-identical code differing only in security-critical logic.\;
\textbf{Gap}: statistical quantification of the semantic gap between paired samples (e.g., CodeBLEU stratification with significance testing).\;
\textbf{Tax.}: LLM-assisted taxonomy that demystifies the root causes of reasoning failures.
\end{table}

To make our positioning explicit, Table~\ref{tab:positioning} compares our study with representative prior work along six axes that we consider essential for diagnosing the reliability of LLM-based vulnerability detectors. PrimeVul~\cite{ding2024primevul} and SV-TrustEval-C~\cite{li2025sv} pioneered paired vulnerable--patch evaluation but focused on out-of-the-box prompting rather than fine-tuning. Risse et al.~\cite{risse2024uncovering} extensively analyzed shortcut learning under semantic-preserving perturbations but restricted their study to encoder-based architectures, leaving contemporary decoder-only LLMs uncovered. Weissberg et al.~\cite{weissberg2025llm} linked LLM predictions to surface-level syntactic metrics but neither performed paired evaluation nor probed the fine-tuning process. To the best of our knowledge, no prior study jointly (i) targets fine-tuned decoder-only LLMs under both vanilla and explicit-reasoning paradigms, (ii) statistically quantifies how the semantic gap between vulnerable code and its patch dictates detection performance, and (iii) systematically dissects explicit-reasoning failures into a structured taxonomy. Our work fills this gap.

\subsection{Motivation}

While recent probing studies and benchmarks have collectively highlighted a pervasive lack of logical robustness in LLMs, a critical blind spot remains in the current literature regarding the supervised fine-tuning (SFT) process and LLM reasoning mechanisms. As the software security community increasingly relies on SFT to adapt general-purpose LLMs into domain-specific security experts, fine-tuning is widely treated as a reliable, ``black-box" performance booster. However, it is fundamentally unclear what these models are actually internalizing during this process. Existing studies diagnosing model failures have primarily evaluated out-of-the-box capabilities, zero-shot prompting strategies, or outdated encoder-based architectures. None have systematically investigated whether the contemporary practice of fine-tuning modern decoder-only LLMs, even when augmented with explicit step-by-step reasoning, genuinely helps models internalize the root causes of vulnerabilities.

This unresolved ambiguity urgently motivates our research. If fine-tuning merely encourages a model to exploit statistical shortcuts, such as the surface-level features shared between a vulnerable function and the surrounding functionally similar code, or the magnitude of textual difference between candidate samples, then the resulting high benchmark scores are dangerously illusory. We refer to this risk as the \textit{Semantic Trap}, operationally characterized in §I along three observable symptoms: pairing-sensitive performance on V2P versus V2N, gap-dictated decisions on (vulnerable, patched) pairs, and fragility to semantic-preserving perturbations. Each symptom is directly testable, and our study is designed to probe all three rather than to argue from a single qualitative observation. By investigating how models behave under paired versus unpaired training data, how their accuracy depends on the semantic gap, and how their reasoning fails when explicitly elicited, this work seeks to answer a pivotal question: do fine-tuned LLMs actually learn the causal logic of software vulnerabilities, or are they merely being trapped as sophisticated pattern matchers?

%% file: Sections/3-Study.tex
\section{Study Design}

\subsection{Research Questions}
\label{RQ}

The objective of this empirical study is to evaluate the effectiveness, robustness, and underlying reasoning behavior of LLMs fine-tuned for software vulnerability detection. We organize the study around three research questions (RQs) that form a deliberate causal chain: RQ1 first asks whether the Semantic Trap can be \emph{observed} by contrasting paired and unpaired training/evaluation data and by stressing models with semantic-preserving perturbations; RQ2 then asks how the trap can be \emph{quantified} by relating detection performance to the textual--syntactic distance between vulnerable and patched code; RQ3 finally asks how the trap can be \emph{anatomized} by inspecting the explicit-reasoning traces that fail-mode predictions produce. Together, the three RQs move from existence to magnitude to mechanism.

\textit{RQ1 -- Observing the trap.} If a fine-tuned model truly internalizes vulnerability root causes, its performance should be stable across (i) different training-data compositions (paired V2P versus unpaired V2N), (ii) cross-dataset evaluation, and (iii) semantic-preserving code perturbations. RQ1 establishes a controlled environment for testing these stability properties and corresponds directly to symptoms~(i) and~(iii) of the Semantic Trap defined in §I. \textbf{RQ1:} \textit{``How do training paradigms (with or without explicit reasoning) and data composition affect the detection effectiveness and robustness of fine-tuned LLMs?''}

\textit{RQ2 -- Quantifying the trap.} The (vulnerable, patched) pairs constructed in RQ1 differ in how much code is changed between the two versions, ranging from one-line condition tweaks to substantial logic rewrites. If a model relies on the magnitude of textual modification as a discriminative shortcut, its accuracy should depend systematically on this gap. RQ2 measures this dependency using CodeBLEU-stratified evaluation and Spearman correlation, and corresponds to symptom~(ii) of the Semantic Trap. \textbf{RQ2:} \textit{``How does the semantic gap between vulnerable code and its corresponding patch statistically influence the detection performance across different fine-tuning paradigms?''}

\textit{RQ3 -- Anatomizing the trap.} Even when models are forced to produce explicit step-by-step reasoning, they still misclassify a non-trivial fraction of inputs. RQ3 reads these failed reasoning traces with an LLM-assisted pipeline to build a taxonomy of error types and to expose what specifically goes wrong when the trap is triggered, distinguishing the failure modes underlying false positives from those underlying false negatives. \textbf{RQ3:} \textit{``What are the primary causes of reasoning failures in fine-tuned LLMs, and how do these failure patterns differ between false positives and false negatives?''}

\subsection{Overview of \framework}
\label{sec:overview}

\begin{figure*}[htbp]
\centering
\includegraphics[width=\textwidth]{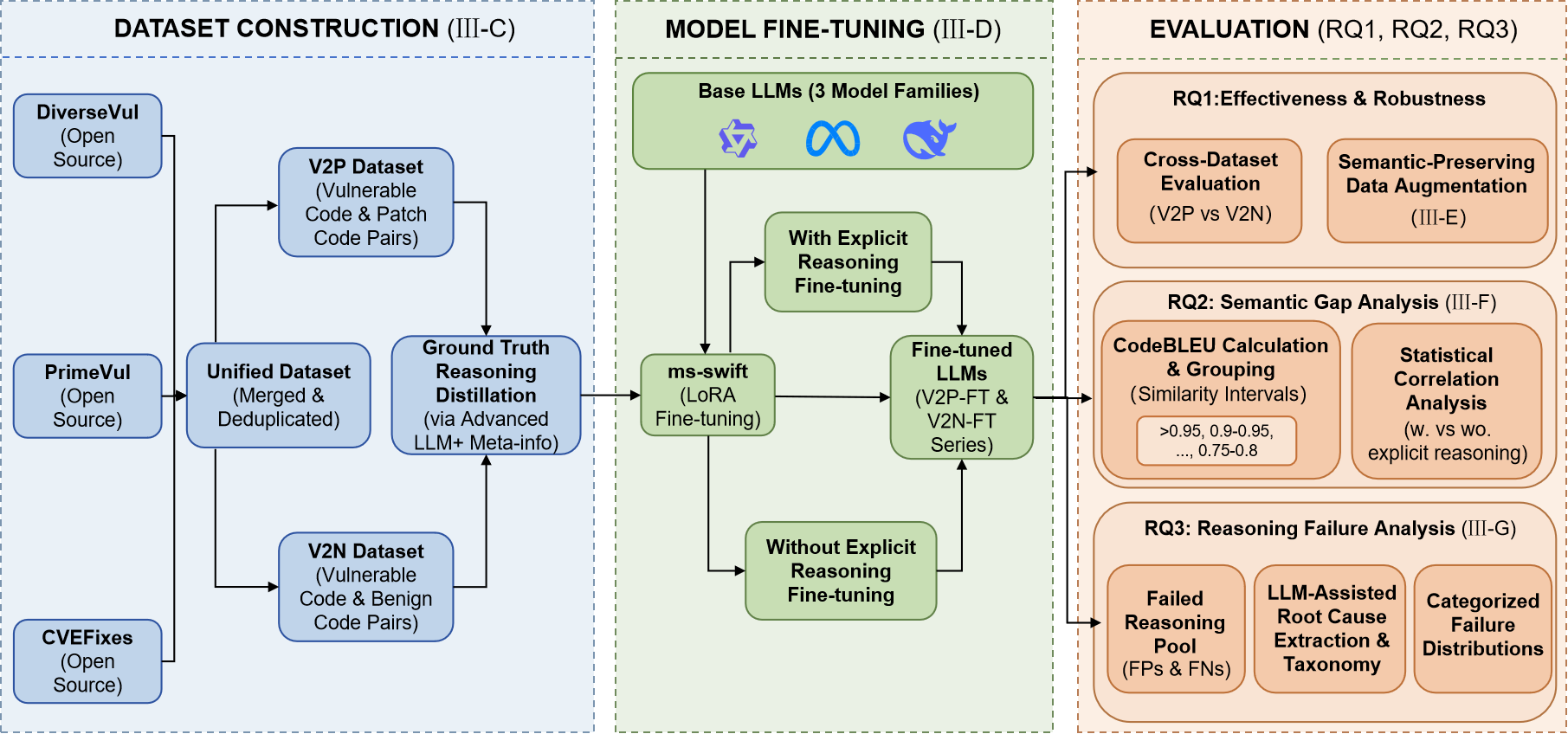}
\caption{Overview of \framework}
\label{fig:overview}
\end{figure*}

To systematically evaluate the performance, robustness, and reasoning capabilities of LLMs in the domain of software vulnerability detection, we design a comprehensive, three-stage evaluation framework, \framework. As illustrated in Figure~\ref{fig:overview}, the framework is structured sequentially into: (1) Dataset Construction, (2) Model Fine-tuning, and (3) Evaluation. This structured approach enables us to address our distinct research questions (RQ1, RQ2, and RQ3) regarding model generalization, sensitivity to semantic nuances, and the fundamental causes of reasoning failures.  
  
The process begins with Stage 1: Dataset Construction (detailed in \S~\ref{sec:dataset}). We consolidate three widely-used open-source vulnerability datasets (namely DiverseVul, PrimeVul, and CVEFixes) into a single unified dataset. To investigate the impact of data composition on model learning, this unified corpus is further structured into two distinct datasets: the \DA, containing pairs of vulnerable code and their corresponding patch code; and the \DB, composed of vulnerable code paired with functionally unrelated non-vulnerable code. We then perform Ground Truth Reasoning Distillation, leveraging an advanced LLM enriched with comprehensive meta-information to generate high-quality, step-by-step reasoning paths for all samples.
  
In Stage 2: Model Fine-tuning (detailed in \S~\ref{sec:finetune}), we select representative base LLMs from three distinct model families to ensure diverse capability baselines. Utilizing the ms-swift~\cite{zhao2024swiftascalablelightweightinfrastructure} framework and the Low-Rank Adaptation (LoRA)~\cite{hu2022lora} technique for parameter-efficient training, we adapt these base models under two distinct training paradigms: with or without explicit reasoning. Applied to both the \DA and \DB training sets, this yields a comprehensive suite of fine-tuned models systematically varying by both training data composition and learning paradigm.  
  
Finally, Stage 3: Evaluation leverages the fine-tuned models and the held-out test sets to systematically investigate our research questions. To address RQ1, we assess both effectiveness and robustness by conducting a cross-dataset evaluation alongside semantic-preserving data augmentations (\S~\ref{sec:augmentation}), comparing performance under standard and perturbed conditions. For RQ2, we conduct a semantic gap analysis (\S~\ref{sec:codebleu}) by grouping \DA test cases based on CodeBLEU similarity intervals, performing statistical correlation analyses to evaluate how varying degrees of semantic divergence influence model predictions under two different paradigms. Lastly, to answer RQ3, we perform an in-depth reasoning failure analysis (\S~\ref{sec:cot_analysis}). By aggregating a pool of failed predictions (False Positives and False Negatives), we utilize an LLM-assisted approach to extract root causes, establish a taxonomy of reasoning errors, and analyze the categorized failure distributions, thereby demystifying the fundamental limitations of current LLMs in vulnerability detection.

\subsection{Dataset Construction}
\label{sec:dataset}
To evaluate the robustness of the fine-tuned large model in vulnerability detection, we propose the \DA and \DB, described in this section.

\textbf{Unified Dataset Merge.} To construct a comprehensive and representative benchmark for vulnerability detection, we integrated three prominent open-source datasets: DiverseVul~\cite{chen2023diversevul}, PrimeVul~\cite{ding2024primevul}, and CVEFixes~\cite{bhandari2021cvefixes}. These datasets were selected based on two primary considerations. First, they are all derived from real-world open-source repositories by crawling vulnerability-fix commits. Consequently, they provide a realistic distribution of code samples, including vulnerable functions, their corresponding fixed versions, and irrelevant non-vulnerable samples. Second, while label noise is a common challenge in commit-based vulnerability datasets, empirical evidence from the PrimeVul study~\cite{ding2024primevul} suggests that all three datasets maintain a label accuracy exceeding 50\%. This ensures a reliable foundation for training and evaluation.

To maintain consistency in programming languages across the unified corpus, we extracted C and C++ samples from CVEFixes, aligning it with the language distribution of the other two sources. During the merging process, we performed hash-based deduplication at the function level. Specifically, for all function samples associated with a given commit, we computed their hash values and retained unique instances. In cases where the same code samples were associated with conflicting labels across different sources, we implemented a priority-based resolution strategy. Specifically, labels from PrimeVul were prioritized, followed by DiverseVul, and finally CVEFixes. This hierarchy is grounded in the reported label accuracy rankings~\cite{ding2024primevul} (PrimeVul \textgreater DiverseVul \textgreater CVEFixes). The final unified dataset comprises a total of 436,489 samples, consisting of 24,758 vulnerable functions, 19,418 fixed versions, and 392,313 irrelevant normal samples.

\textbf{\DA and \DB construction.} Based on the Unified Dataset, we constructed two specialized datasets, namely \DA (Vulnerability-to-Patch) and \DB (Vulnerability-to-Normal), to support model training and evaluation. The construction process was executed at the granularity of individual commit-IDs using the following three steps:
\textit{1) Sample Pairing and Filtering:} Within each commit-ID, we used the function name as a unique fingerprint to identify corresponding code versions. Samples sharing the same fingerprint but possessing opposite labels (vulnerable vs. fixed) were identified as a vulnerability-fix pair. Conversely, samples labeled as ``normal'' whose fingerprints did not match any vulnerable counterparts within the same commit were categorized as irrelevant normal samples.
\textit{2) Dataset Composition:} The \DA was formed by directly pairing vulnerable samples with their corresponding fixed versions. To construct the \DB, we reused the \emph{same} set of vulnerable samples and paired them with an equal number of irrelevant normal samples randomly selected from the previously filtered ``normal'' pool. By design, the vulnerable side of \DAShort and \DBShort is fully overlapping (24{,}747 vulnerable functions in both), while the non-vulnerable side is disjoint: \DAShort uses 19{,}418 patched versions, whereas \DBShort uses 19{,}418 unrelated normal functions. This shared-vulnerability construction is intentional: it lets cross-dataset comparisons (e.g., V2N$\rightarrow$V2P in RQ1) attribute any performance change to the \emph{negative} side, the choice between a patched counterpart and an unrelated benign sample, rather than to a shift in the vulnerable population.
\textit{3) Data Partitioning:} Both \DA and \DB were partitioned into training and test sets using a 9:1 ratio at the commit level, so that a vulnerable function and its patched/normal counterpart are always assigned to the same split (no cross-split leakage of a (vuln, fix) pair).

\textbf{Ground Truth Reasoning Distillation.} To support model fine-tuning and evaluation under the explicit-reasoning paradigm, we performed a reasoning distillation process for every sample in both the \DAShort and \DBShort using an advanced LLM (e.g., GPT-5.4). Rather than relying solely on the code snippet, we enriched the distillation process by providing the LLM with comprehensive oracle-level context. Specifically, the input to the LLM included the source code, its definitive true label, and extensive meta-information: the corresponding CWE ID, the official description of that CWE, the originating project repository, and the developer's commit message. Through a carefully designed prompt, we instructed the LLM to synthesize all these known facts to generate a logical, step-by-step reasoning path that explains exactly whether and why the code constitutes a vulnerability. The resulting high-quality analytical explanations were systematically collected and treated as the ``ground truth reasoning'' to supervise the fine-tuning process.

\subsection{LLM Finetuning}
\label{sec:finetune}

\textbf{Motivation and Model Selection.} To adapt LLMs to the specialized domain of software vulnerability detection, prior research has extensively leveraged vulnerability-related datasets for supervised fine-tuning. In this study, we incorporate the fine-tuning process into \framework to assess the robustness and adaptability of state-of-the-art LLMs in identifying code vulnerabilities. Specifically, we select three of the most prominent open-source model families: Qwen, Llama, and DeepSeek. These models are fine-tuned on the training sets of our constructed \DA and \DB to establish a baseline for subsequent evaluation and analysis.

\textbf{Fine-tuning Framework: ms-swift.} The fine-tuning process is implemented using ms-swift~\cite{zhao2024swiftascalablelightweightinfrastructure}, a comprehensive framework for large-scale and multimodal model lifecycle management provided by the ModelScope community. 
By leveraging ms-swift, we ensure a high-performance and scalable pipeline from model training to quantization and deployment.

\textbf{Parameter-Efficient Fine-Tuning with LoRA.} To achieve lightweight adaptation while mitigating the risk of catastrophic forgetting, a common issue where full-parameter fine-tuning leads to the loss of pre-trained general knowledge, we employ Low-Rank Adaptation (LoRA)~\cite{hu2022lora}, the same fine-tuning method as previous works~\cite{boi2024smart, ma2024combining, du2024generalization}. The core philosophy of LoRA is to freeze the original weights of the Pre-trained Language Model (PLM) and introduce trainable low-rank decomposition matrices as a bypass. Given a pre-trained weight matrix $W_0\in\mathbb{R}^{d\times k}$, its update during fine-tuning is represented as:
\[W_0+\Delta W=W_0+BA\]
where $B\in\mathbb{R}^{d\times r}$ and $A\in\mathbb{R}^{r\times k}$ are the low-rank matrices, with the rank $r\ll\min(d,k)$. This approach significantly reduces the number of trainable parameters, thereby decreasing the computational overhead and memory footprint.

\subsection{Data Augmentation}
\label{sec:augmentation}
As part of our investigation in RQ1 to assess the robustness of fine-tuned LLMs in vulnerability detection, we perform data augmentation on the test partitions of the V2P and V2N datasets. The primary objective is to evaluate whether the models can maintain consistent performance when confronted with functional pattern variations that preserve the underlying program logic.

We employ \textbf{Semantic-Preserving Data Augmentation}, a technique that introduces perturbations into the source code while strictly adhering to the original program's execution logic and semantics~\cite{ullah2024llms}. This ensures that any identified vulnerability or "fixed" state remains unchanged despite the structural or lexical modifications. Specifically, we implement the same seven distinct transformation methods:

\begin{itemize}[]
    \item \textbf{Random Parameter Renaming}: Replacing function parameter identifiers with randomly generated strings.
    \item \textbf{Random Function Renaming}: Assigning a new, random name to the function while updating its internal references.
    \item \textbf{Unreachable Code Injection}: Inserting code blocks that are logically impossible to execute (e.g., within an if (0) block), thereby testing the model's ability to ignore dead code.
    \item \textbf{Comment-based Perturbation}: Adding random snippets of code or text within comment delimiters (/* ... */ or //), which should be ignored by the model's semantic analysis.
    \item \textbf{Whitespace Insertion}: Injecting redundant spaces or tabs between tokens to alter the physical layout of the code.
    \item \textbf{Redundant Function Addition}: Appending an auxiliary, unused function to the code sample, increasing the input complexity without affecting the target function.
    \item \textbf{Line-Break Insertion}: Adding multiple next-line characters (\textbackslash n) between statements to perturb the sequence of code tokens.
\end{itemize}

By applying these transformations, we generate an augmented version of the test suite. Specifically, for each original test case, we generated seven distinct variants, with each variant corresponding to one of the seven transformation strategies. This resulted in a 7x expansion of the test set, ensuring a comprehensive evaluation across different types of semantic-preserving perturbations. The fine-tuned LLMs (Qwen, Llama, and DeepSeek) are subsequently evaluated on this augmented set to measure the degradation or stability of their detection capabilities.

\subsection{Semantic Gap Analysis}
\label{sec:codebleu}
To address RQ2, it is essential to quantify the semantic distance between vulnerable functions and their corresponding patches. While vulnerability detection often hinges on subtle code modifications, the magnitude of these changes varies significantly across different samples. To provide an objective measurement of this "semantic gap," we adopt CodeBLEU~\cite{ren2020codebleu} as our primary evaluation metric for each (vulnerability, patch) pair.

Originally proposed for evaluating code generation quality, CodeBLEU extends the traditional BLEU score by incorporating syntactic and semantic structures inherent to source code. Specifically, it augments n-gram matching with (1) abstract syntax tree (AST) similarity, which captures structural alignment, and (2) data-flow alignment, which reflects semantic equivalence in variable usage and program behavior. By integrating these program-aware features, CodeBLEU provides a more faithful and objective assessment of code similarity from a programming semantics perspective, rather than relying solely on lexical overlap, making it particularly suitable for measuring the semantic distance between a vulnerability and its corresponding fix.

In our implementation, we calculate a CodeBLEU score for every (vulnerability, patch) pair within the V2P test set. A higher score indicates a smaller semantic gap (i.e., the patch is more similar to the original vulnerable code), whereas a lower score suggests more extensive modifications. To systematically investigate how the degree of code modification affects model performance, we categorized the test samples into different groups by a fixed internal of 0.05.
The fine-tuned LLMs (Qwen, Llama, and DeepSeek) were evaluated across these intervals to observe performance fluctuations relative to the semantic gap. 

\subsection{LLM-Assisted Wrong Reasoning Analysis}
\label{sec:cot_analysis}

\begin{figure}[htbp]
\centering
\includegraphics[width=\columnwidth]{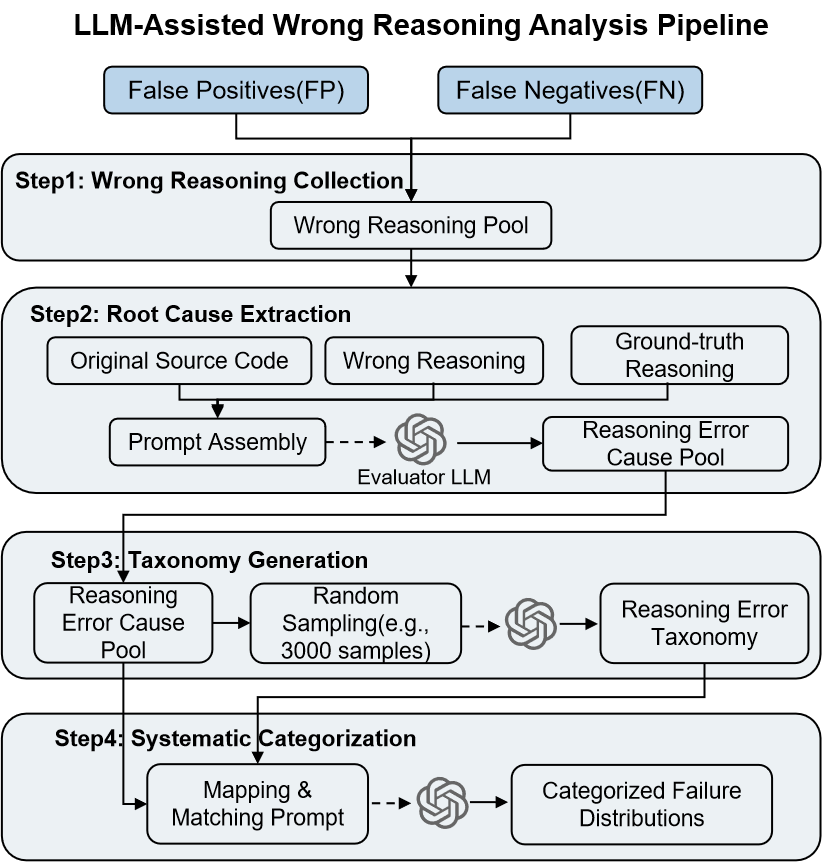}
\caption{The Pipeline of LLM-Assisted Wrong Reasoning Analysis}
\label{fig:cot_analysis}
\end{figure}

To address RQ3, it is essential to conduct an in-depth investigation into the incorrect reasoning generated by the fine-tuned models. Relying solely on quantitative metrics is insufficient to reveal the fundamental logical flaws or semantic blind spots of LLMs. Therefore, we designed a four-step LLM-assisted error analysis pipeline to systematically extract, categorize, and analyze the underlying reasons for reasoning errors.   

As shown in Figure~\ref{fig:cot_analysis}, We begin the pipeline by systematically aggregating all failure cases, specifically, the False Positives (FPs) and False Negatives (FNs) generated by the fine-tuned models across all experimental configurations to form a comprehensive wrong reasoning pool. For each sample within this pool, we construct a carefully designed prompt that juxtaposes the original source code snippet, the model's wrong reasoning, and the ground-truth reasoning (distilled in \S\ref{sec:dataset}). We then employ an advanced evaluator LLM to comparatively analyze these inputs and briefly summarize the essential root cause of the reasoning derailment, thereby populating a reasoning error cause pool. To derive a structured understanding of these diverse failures, we randomly sample a substantial subset (e.g., 3,000 instances) from this error cause pool. This subset is fed back into the evaluator LLM with the instruction to synthesize, cluster, and abstract the specific error descriptions into a standardized taxonomy of reasoning error categories. Finally, to quantify the prevalence of these specific blind spots, we map all individual failures to the established taxonomy. By prompting the evaluator LLM with each sample's specific error cause summary alongside the formalized taxonomy, every failure is classified into its single most appropriate error category, ultimately enabling the detailed quantitative distribution analysis presented in our findings.  

%% file: Sections/4-Experiments.tex
\section{Experiments}

In this section, we will first introduce the setup about the expriments, and then analyze the experiment results to answer the following three research question.

\begin{itemize}[]
\item \textbf{RQ1:} How do training paradigms (with or without explicit reasoning) and data composition affect the detection effectiveness and robustness of fine-tuned LLMs?
\item \textbf{RQ2:} How does the semantic gap between vulnerable code and its corresponding patch statistically influence the detection performance across different fine-tuning paradigms?
\item \textbf{RQ3:} What are the primary causes of reasoning failures in fine-tuned LLMs, and how do these failure patterns differ between false positives and false negatives?
\end{itemize}

\subsection{Experimental Setup}

\begin{figure}
\begin{tcolorbox}[title=Prompt Template]
\small
  \textbf{User:}\\
Below is an instruction that describes a classification task.
Devise a label name suitable for categorizing items as either vulnerable or safe.

\#\#\# Instruction:
Please review the code. Please find out if it is vulnerable. \red{Please think step by step (Chain of Thought).}

\#\#\# Input: \\
  \textasciigrave\textasciigrave\textasciigrave C++\\
  \blue{\{code\}}\\
  \textasciigrave\textasciigrave\textasciigrave \\
\red{\#\#\# Output Format:
Please output the result in the following JSON format: \\
\{\{"reason": "CoT reason", "result": "The code is safe/vulnerable"\}\}}

\#\#\# Response:
\end{tcolorbox}
\caption{Prompt Template during fine-tuning and evaluation.}
\label{fig:promptTemplate}
\end{figure}

This section details the experimental design used to evaluate fine-tuned LLMs for vulnerability detection, including the selection of datasets, model configurations, training protocols, and the hardware environment utilized for the study.

To do a comprehensive analysis, we choose 5 different state-of-the-art open-source large languages models for evaluation: Qwen3-8B~\cite{qwen3technicalreport}, Qwen2.5-Coder-7B-Instruct~\cite{hui2024qwen2}, Llama3.1-8B~\cite{grattafiori2024llama}, DeepSeek-LLM-7B-Chat~\cite{deepseekai2024deepseekv3technicalreport}, and DeepSeek-Coder-6.7B-Instruct~\cite{deepseek-coder}. 
Notably, Qwen2.5-Coder-7B-Instruct and DeepSeek-Coder-6.7B-Instruct are instruction-tuned models explicitly optimized for code generation, enabling us to investigate the effect of domain-specific fine-tuning. 
All models fall within the 7B–8B parameter range, offering a favorable trade-off between capability and computational efficiency for fine-tuning.

Each model was fine-tuned under two distinct training paradigms. For both paradigms, the models were trained on \DA ~(paired vulnerable code with patched version) and \DB ~(non-paired cases) (\S\ref{sec:dataset}) respectively, utilizing the ms-swift~\cite{zhao2024swiftascalablelightweightinfrastructure} framework with LoRA (Low-Rank Adaptation)~\cite{hu2022lora}.
The LoRA is trained with a \textit{rank} of 8 and an \textit{alpha value} of 32, with all other parameters remaining default, training for 3 epochs.
A consistent prompt template depicted in Figure~\ref{fig:promptTemplate} was applied in all training and models to standardize the input-output structure, where the text highlighted in red indicates the specific differences between two fine-tuning paradigms.

Furthermore, to conduct an in-depth investigation into the causes of reasoning failures (addressed in RQ3), we incorporated an LLM-assisted analysis pipeline. Specifically, we utilized the Qwen3.5-plus model as the backbone engine to systematically extract, categorize, and analyze the underlying reasons for reasoning errors across both false positive and false negative samples.

All experiments, including both fine-tuning and inference, were conducted on 8 NVIDIA A100 GPUs. 
For inference, we used the vLLM~\cite{kwon2023efficient} inference engine with the temperature set to 0, while other parameters set to default, to ensure deterministic generation and reproducible results.
To comprehensively assess model performance across all experimental settings, we treat the detection task as a binary classification task, employing five standard metrics: accuracy (Acc), precision (Pre), recall (Rec), F1-score (F1), and false positive rate (FPR). 

\subsection{Effectiveness and Robustness Analysis: The Impact of Training Paradigms and Data Composition (RQ1)}

\begin{table}[t]
\centering
\caption{Pre- vs.\ post-SFT performance of Qwen3-8B on the V2P and V2N test sets. Qwen3-8B is selected here because its native ``think'' mode lets the base model produce explicit reasoning \emph{without} fine-tuning, providing the strongest control: any improvement after SFT cannot be attributed merely to enabling reasoning. The remaining four evaluated models do not provide a comparable native reasoning mode, and their pre-SFT zero-shot behavior on similar benchmarks is already well documented in prior work (e.g., \cite{ding2024primevul, li2025sv}); we therefore do not repeat those measurements and focus our pre-vs-post SFT contrast on Qwen3-8B.}
\label{table:finding1_pre}
\small
\setlength{\tabcolsep}{4pt}
\renewcommand{\arraystretch}{1.15}
\begin{tabular}{cc|ccccc}
\hline
\textbf{Eval.} & \textbf{Setting} & \textbf{acc} & \textbf{pre} & \textbf{rec} & \textbf{f1} & \textbf{fpr} \\ \hline
\multirow{2}{*}{V2P} & SFT     & 52.62 & 60.49 & 49.30 & 54.33 & 42.95 \\
                    & non-SFT & 48.25 & 60.39 & 28.18 & 38.43 & 24.81 \\
\multirow{2}{*}{V2N} & SFT     & 66.43 & 75.04 & 61.22 & 67.43 & 26.73 \\
                    & non-SFT & 50.77 & 62.71 & 27.18 & 37.92 & 20.01 \\ \hline
\end{tabular}
\end{table}

\begin{table*}[]
\centering
\caption{Performance comparison of LLMs fine-tuned on vulnerability detection under cross-dataset evaluation settings. The row labels "w." and "wo." denote models fine-tuned with and without explicit reasoning, respectively. The columns labeled \DAShort-\textgreater\DAShort ~and \DBShort-\textgreater\DBShort ~denote models fine-tuned and evaluated on the same dataset (\DA or \DB, respectively), while \DAShort-\textgreater\DBShort ~and \DBShort-\textgreater\DAShort ~represent cross-dataset evaluations. Metrics include accuracy (acc), precision (pre), recall (rec), F1-score (f1), and false positive rate (fpr).}
\begin{tabular}{cc|lllll|lllll}
\hline
\multicolumn{2}{c|}{\multirow{2}{*}{\textbf{Models}}}& \multicolumn{5}{c|}{\textbf{\DAShort-\textgreater\DAShort}} & \multicolumn{5}{c}{\textbf{\DAShort-\textgreater\DBShort}} \\
\multicolumn{2}{c|}{} & \multicolumn{1}{c}{acc} & \multicolumn{1}{c}{pre} & \multicolumn{1}{c}{rec} & \multicolumn{1}{c}{f1} & \multicolumn{1}{c|}{fpr} & \multicolumn{1}{c}{acc} & \multicolumn{1}{c}{pre} & \multicolumn{1}{c}{rec} & \multicolumn{1}{c}{f1} & \multicolumn{1}{c}{fpr} \\ \hline
\multirow{2}{*}{\textbf{Qwen3-8B}}& \textbf{wo.}& 57.47& 60.74& 72.41& 66.06& 62.48 & 54.62& 57.48& 77.06& 65.85& 74.86\\
& \textbf{w.}& 52.62& 60.49& 49.30& 54.33& 42.95 & 50.98& 57.81& 50.34& 53.82& 48.19\\
\multirow{2}{*}{\textbf{Qwen2.5-Coder-7B-Instruct}}& \textbf{wo.}& 58.08& 61.29& 72.41& 66.39& 61.05 & 55.25& 57.44& 81.73& 67.47& 79.53\\
& \textbf{w.}& 53.36& 60.32& 53.87& 56.92& 47.32 & 51.67& 57.95& 54.00& 55.91& 51.39\\
\multirow{2}{*}{\textbf{Llama-3.1-8B-Instruct}}& \textbf{wo.}& 58.26& 60.52& 77.65& 68.02& 67.62 & 54.16& 56.61& 82.49& 67.14& 83.04\\
& \textbf{w.}& 53.93& 59.67& 59.84& 59.75& 53.95 & 53.48& 58.65& 61.02& 59.81& 56.42\\
\multirow{2}{*}{\textbf{deepseek-llm-7b-chat}} & \textbf{wo.}& 55.64& 59.09& 73.45& 65.49& 68.28 & 54.11& 55.91& 80.29& 65.92& 78.23\\
& \textbf{w.}& 51.85& 61.07& 46.31& 52.67& 40.54 & 49.43& 50.26& 46.47& 48.29& 47.50\\
\multirow{2}{*}{\textbf{deepseek-coder-6.7b-instruct}} & \textbf{wo.}& 55.63& 61.14& 56.73& 58.85& 45.75 & 51.82& 64.29& 59.61& 61.86& 63.00\\
& \textbf{w.}& 51.74& 58.62& 52.80& 55.56& 49.67 & 51.43& 57.30& 53.94& 55.57& 51.81\\ \hline
\multirow{2}{*}{\textbf{Models}}  & \multicolumn{1}{l|}{} & \multicolumn{5}{c|}{\textbf{\DBShort-\textgreater\DAShort}} & \multicolumn{5}{c}{\textbf{\DBShort-\textgreater\DBShort}} \\
& \multicolumn{1}{l|}{} & \multicolumn{1}{c}{acc} & \multicolumn{1}{c}{pre} & \multicolumn{1}{c}{rec} & \multicolumn{1}{c}{f1} & \multicolumn{1}{c|}{fpr} & \multicolumn{1}{c}{acc} & \multicolumn{1}{c}{pre} & \multicolumn{1}{c}{rec} & \multicolumn{1}{c}{f1} & \multicolumn{1}{c}{fpr} \\ \hline
\multirow{2}{*}{\textbf{Qwen3-8B}}& \textbf{wo.}& 58.13& 57.96& 97.46& 72.69& 94.38 & 81.28& 82.95& 84.37& 83.65& 22.78\\
& \textbf{w.}& 53.51& 58.62& 63.42& 60.92& 59.70 & 66.43& 75.04& 61.22& 67.43& 26.73\\
\multirow{2}{*}{\textbf{Qwen2.5-Coder-7B-Instruct}}& \textbf{wo.}& 58.72& 58.04& 97.82& 72.86& 94.43 & 81.63& 83.63& 84.13& 83.88& 21.65\\
& \textbf{w.}& 53.44& 58.46& 64.15& 61.17& 60.85 & 67.19& 76.04& 61.58& 68.05& 25.46\\
\multirow{2}{*}{\textbf{Llama-3.1-8B-Instruct}}& \textbf{wo.}& 57.85& 57.84& 96.98& 72.46& 94.38 & 81.23& 82.50& 84.97& 83.72& 23.67\\
& \textbf{w.}& 54.81& 58.82& 69.41& 63.67& 64.60 & 68.36& 74.47& 67.23& 70.66& 30.16\\
\multirow{2}{*}{\textbf{deepseek-llm-7b-chat}} & \textbf{wo.}& 57.61& 57.76& 96.94& 72.39& 95.21 & 75.36& 76.56& 79.86& 78.18& 30.21\\
& \textbf{w.}& 51.90& 61.14& 46.31& 52.70& 40.44 & 61.92& 69.06& 45.44& 54.81& 21.04\\
\multirow{2}{*}{\textbf{deepseek-coder-6.7b-instruct}} & \textbf{wo.}& 58.01& 57.95& 97.37& 72.65& 94.79 & 79.74& 80.73& 83.55& 82.12& 25.05\\
& \textbf{w.}& 53.32& 59.04& 58.51& 58.78& 53.51 & 64.33& 74.14& 55.75& 63.64& 24.75\\ \hline
\end{tabular}
\label{table:finding1}
\end{table*}

\begin{figure}[htbp]
\centering
\includegraphics[width=\columnwidth]{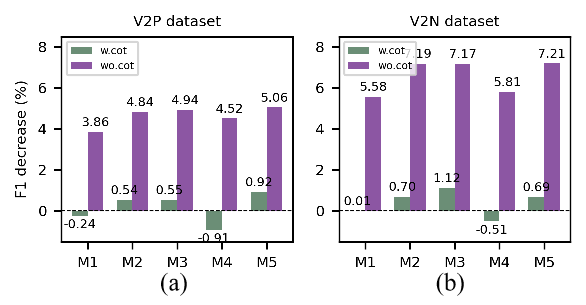}
\caption{F1-score decrease after applying data augmentation to the test sets. The bar charts compare the performance degradation (in percentage points) of five LLMs fine-tuned with and without explicit reasoning. (a) Evaluation results on the augmented \DAShort test set. (b) Evaluation results on the augmented \DBShort test set. M1 to M5 represent Qwen3-8B, Qwen2.5-Coder-7B-Instruct, Llama-3.1-8B-Instruct, deepseek-llm-7b-chat, and deepseek-coder-6.7b-instruct, respectively.}
\label{fig:rq1}
\end{figure}

In this RQ, we evaluate the vulnerability detection capabilities of LLMs under different training paradigms (with and without explicit reasoning) and investigate how the composition of training data influences their performance.
To this end, we fine-tuned five representative LLMs under the two paradigms on two distinct datasets: \DA and \DB (\S~\ref{sec:dataset}).
Before interpreting any cross-dataset or robustness results, we verify the basic premise that SFT does change model behavior in a non-trivial way, focusing on the most conservative control: Qwen3-8B, the only evaluated model with a native ``think'' mode that lets the base model already produce explicit reasoning without fine-tuning. Table~\ref{table:finding1_pre} reports its pre- vs.\ post-SFT performance on both V2P and V2N. SFT lifts V2N F1 from 37.92\% to 67.43\% (+29.5 pp) and V2P F1 from 38.43\% to 54.33\% (+15.9 pp), confirming that the improvement after SFT cannot be attributed merely to enabling reasoning, the base model is already reasoning, yet it is materially worse. The remaining four evaluated models do not provide a comparable native reasoning mode, and their pre-SFT zero-shot behavior is already well documented by prior benchmarks~\cite{ding2024primevul, li2025sv}; we therefore do not duplicate those measurements here. With this premise established for Qwen3-8B, the remainder of RQ1 focuses on \emph{how} SFT shifts the model: whether it teaches genuine security logic, or whether the improvement is concentrated on the unpaired V2N setting while collapsing under V2P, perturbations, and cross-dataset transfer.
The comprehensive results for the standard cross-dataset evaluation and the data augmentation tests are presented in Table~\ref{table:finding1} and Figure~\ref{fig:rq1}, respectively.

As shown in Table~\ref{table:finding1}, regardless of the fine-tuning paradigm (with or without explicit reasoning), models exhibit significantly different performance depending on the composition of the training data.
Specifically, the experimental results of the \DBShort-to-\DBShort configuration are substantially better than those of the \DAShort-to-\DAShort configuration.
For instance, fine-tuning without explicit reasoning (wo.), \texttt{Qwen3-8B} achieves an F1-score of 83.65\% on \DBShort-to-\DBShort, but only 66.06\% on \DAShort-to-\DAShort. Similarly, \texttt{Qwen2.5-Coder-7B-Instruct} drops from 83.88\% (\DBShort-to-\DBShort) to 66.39\% (\DAShort-to-\DAShort). 
For finetuning with explicit reasoning (w.), \texttt{Llama-3.1-8B-Instruct} drops from 70.66\% (\DBShort-to-\DBShort) to 59.75\% (\DAShort-to-\DAShort).
Furthermore, this performance gap persists in cross-dataset evaluations, where models trained on \DB exhibit noticeably better results than those trained on \DA. For example, the F1-score of (wo.) is 72.39\% in the \DBShort-to-\DAShort setting, noticeably higher than its 65.92\% in the \DAShort-to-\DBShort setting. A similar trend is observed under another paradigm, where \texttt{deepseek-coder-6.7b-instruct} (w.) scores 58.78\% on \DBShort-to-\DAShort compared to just 55.57\% on \DAShort-to-\DBShort.

This stark contrast highlights the profound impact of dataset construction. In \DA, the positive and negative samples (vulnerable code and its corresponding patched version) share nearly identical functional semantics and structures. This proximity makes it extremely difficult for models to distinguish between them, resulting in poor training performance. Conversely, the seemingly superior metrics in \DB suggest that models might merely be capturing superficial code patterns or broad functional semantics as a "shortcut" for learning, rather than identifying the intrinsic logical flaws.

\finding{}{
The composition of the training dataset significantly impacts the model's performance. Models struggle to learn from \DA (paired vulnerable and patched code), indicating that high performance on \DB is likely an artifact of models adopting superficial code patterns as learning shortcuts rather than capturing the actual root causes of vulnerabilities.
}

To further investigate the learning behaviors, we evaluate the robustness of models fine-tuned with (w.) and without (wo.) explicit reasoning.
As evident in the \DBShort-to-\DAShort setting, models trained without explicit reasoning exhibit a prohibitively high False Positive Rate (FPR) alongside exceedingly high recall. For example, the recall for \texttt{Llama-3.1-8B-Instruct} (wo.) reaches 96.98\%, but its FPR surges to 94.38\%. This indicates that models finetuned without explicit reasoning overfit to the surface patterns learned from \DB; when presented with a patched version in \DA, they fail to recognize the fix and blindly flag it as vulnerable simply because it shares the same functional pattern.
In contrast, fine-tuning with explicit reasoning significantly mitigates this issue, reducing the FPR of \texttt{Llama-3.1-8B-Instruct} from 94.38\% to 64.60\%.

Furthermore, the data augmentation experiments shown in Figure~\ref{fig:rq1} provide additional evidence of robustness differences. When evaluated on perturbed test sets, the F1-scores of models fine-tuned without explicit reasoning drop by between 3.86 and 7.21 percentage points. By contrast, models fine-tuned with explicit reasoning remain comparatively stable, with F1 decreases largely contained within 1.12 percentage points, and even showing slight improvements in certain cases (e.g., $-0.91\%$ for \texttt{deepseek-llm-7b-chat} on the \DA dataset).

\finding{}{
Models fine-tuned without explicit reasoning tend to overfit to the surface patterns of vulnerable code and ignore the true root causes, leading to massive false alarms on patched code. Incorporating fine-tuning with explicit reasoning forces the model to reason, making it significantly more robust against distribution shifts and data perturbations.
}

Despite the evident benefits in robustness, the overall performance metrics for with explicit reasoning paradigm are less than satisfactory.
Across almost all experimental groups in Table~\ref{table:finding1}, the F1-scores of the models fine-tuned with explicit reasoning are consistently lower than those of the models fine-tuned without explicit reasoning.
This performance degradation is primarily driven by a substantial drop in recall, meaning models fine-tuned with explicit reasoning produce considerably more False Negatives (FN).
For instance, in the \DAShort-to-\DAShort setting, the recall for \texttt{deepseek-llm-7b-chat} drops drastically from 73.45\% (wo.) to 46.31\% (w.), leading to an F1-score decrease from 65.49\% to 52.67\%.
The requirement to explicitly generate a reasoning chain makes models more conservative. If a model fails to construct a complete and correct logical path to identify the subtle flaw, it defaults to predicting the code as benign, thereby missing the vulnerability.

\finding{}{
While fine-tuning with explicit reasoning successfully enhances model robustness and mitigates superficial overfitting, it comes at the cost of overall model performance. The strict reasoning requirements introduce more False Negatives, resulting in significantly lower recall and F1-scores compared to the paradigm without explicit reasoning.
}

\subsection{Semantic Gap Analysis: Statistical Correlation Between Patch Proximity and Model Performance (RQ2)}

\begin{table*}
\centering
\caption{Performance of LLMs fine-tuned on restructured \DA, evaluated across vulnerability–patch pairs stratified by CodeBLEU score. The table compares two fine-tuning paradigms: with and without explicit reasoning. The ‘count’ column indicates the number of samples within each interval. In the 'relevance' column, 'sc' denotes the Spearman's correlation coefficient, the brackets indicate the 95\% confidence interval, and 'p' represents the p-value.}
\begin{tabular}{cclllllcc}
\hline
\multicolumn{9}{c}{\textbf{without explicit reasoning}}\\
\textbf{model}& \textbf{CodeBleu\_Score} & \multicolumn{1}{c}{\textbf{acc}} & \multicolumn{1}{c}{\textbf{pre}} & \multicolumn{1}{c}{\textbf{rec}} & \multicolumn{1}{c}{\textbf{f1}} & \multicolumn{1}{c}{\textbf{fpr}} & \textbf{count} & \textbf{relevance}\\ \hline
\multirow{5}{*}{\textbf{Qwen3-8B}}   & \multicolumn{1}{c|}{\textbf{\textgreater{}0.95}} & 53.79& 54.38& 46.98& 50.41& 39.41   & 1426& \multirow{5}{*}{\begin{tabular}[c]{@{}c@{}}sc:-0.0744\\ {[}-0.1079,-0.0428{]}\\ p:3.4084e-6\end{tabular}} \\
 & \multicolumn{1}{c|}{\textbf{0.9-0.95}}& 57.35& 59.14& 47.55& 52.72& 32.85   & 694&\\
 & \multicolumn{1}{c|}{\textbf{0.85-0.9}}& 57.77& 58.40& 54.06& 56.15& 38.52   & 566&\\
 & \multicolumn{1}{c|}{\textbf{0.8-0.85}}& 60.65& 61.69& 56.21& 58.82& 34.91   & 338&\\
 & \multicolumn{1}{c|}{\textbf{0.75-0.8}}& 64.11& 64.96& 61.29& 63.07& 33.06   & 248&\\ \hline
\multirow{5}{*}{\textbf{Qwen2.5-Coder-7B-Instruct}}& \multicolumn{1}{c|}{\textbf{\textgreater{}0.95}} & 54.49& 55.37& 46.35& 50.46& 37.36   & 1424& \multirow{5}{*}{\begin{tabular}[c]{@{}c@{}}sc:-0.0721\\ {[}-0.1024,-0.0397{]}\\ p:6.8837e-6\end{tabular}} \\
 & \multicolumn{1}{c|}{\textbf{0.9-0.95}}& 58.76& 60.41& 50.86& 55.23& 33.33   & 696&\\
 & \multicolumn{1}{c|}{\textbf{0.85-0.9}}& 60.35& 60.81& 58.25& 59.50& 37.54   & 570&\\
 & \multicolumn{1}{c|}{\textbf{0.8-0.85}}& 60.71& 61.54& 57.14& 59.26& 35.71   & 336&\\
 & \multicolumn{1}{c|}{\textbf{0.75-0.8}}& 63.31& 64.86& 58.06& 61.28& 31.45   & 248&\\ \hline
\multirow{5}{*}{\textbf{Llama-3.1-8B-Instruct}}  & \multicolumn{1}{c|}{\textbf{\textgreater{}0.95}} & 53.79& 55.25& 39.83& 46.29& 32.26   & 1426& \multirow{5}{*}{\begin{tabular}[c]{@{}c@{}}sc:-0.048\\ {[}-0.0801,-0.0157{]}\\ p:2.7695e-3\end{tabular}}  \\
 & \multicolumn{1}{c|}{\textbf{0.9-0.95}}& 56.88& 59.02& 44.99& 51.06& 31.23   & 698&\\
 & \multicolumn{1}{c|}{\textbf{0.85-0.9}}& 55.40& 57.21& 42.81& 48.97& 32.01   & 556&\\
 & \multicolumn{1}{c|}{\textbf{0.8-0.85}}& 58.67& 60.14& 51.45& 55.45& 34.10   & 346&\\
 & \multicolumn{1}{c|}{\textbf{0.75-0.8}}& 60.00& 61.06& 55.20& 57.98& 35.20   & 250&\\ \hline
\multirow{5}{*}{\textbf{deepseek-llm-7b-chat}}   & \multicolumn{1}{c|}{\textbf{\textgreater{}0.95}} & 50.74& 51.68& 23.95& 32.73& 22.43   & 1285& \multirow{5}{*}{\begin{tabular}[c]{@{}c@{}}sc:-0.0335\\ {[}-0.0658,-0.0015{]}\\ p:4.1954e-2\end{tabular}} \\
 & \multicolumn{1}{c|}{\textbf{0.9-0.95}}& 53.89& 56.07& 35.93& 43.80& 28.14   & 668&\\
 & \multicolumn{1}{c|}{\textbf{0.85-0.9}}& 52.43& 53.10& 43.17& 47.62& 38.27   & 555&\\
 & \multicolumn{1}{c|}{\textbf{0.8-0.85}}& 53.96& 54.86& 47.88& 51.13& 39.88   & 328&\\
 & \multicolumn{1}{c|}{\textbf{0.75-0.8}}& 58.47& 59.29& 54.03& 56.54& 37.10   & 248&\\ \hline
\multirow{5}{*}{\textbf{deepseek-coder-6.7b-instruct}} & \multicolumn{1}{c|}{\textbf{\textgreater{}0.95}} & 52.99& 54.90& 33.90& 41.91& 27.89   & 1421& \multirow{5}{*}{\begin{tabular}[c]{@{}c@{}}sc:-0.0666\\ {[}-0.0973,-0.0377{]}\\ p:3.3016e-5\end{tabular}} \\
 & \multicolumn{1}{c|}{\textbf{0.9-0.95}}& 55.00& 58.37& 34.86& 43.65& 24.86   & 700&\\
 & \multicolumn{1}{c|}{\textbf{0.85-0.9}}& 57.14& 61.49& 38.21& 47.14& 23.93   & 560&\\
 & \multicolumn{1}{c|}{\textbf{0.8-0.85}}& 58.53& 62.39& 42.94& 50.87& 25.88   & 340&\\
 & \multicolumn{1}{c|}{\textbf{0.75-0.8}}& 62.40& 65.05& 53.60& 58.77& 28.80   & 250&\\ \hline
\multicolumn{9}{c}{\textbf{with explicit reasoning}}\\
\textbf{model}& \textbf{CodeBleu\_Score} & \multicolumn{1}{c}{\textbf{acc}} & \multicolumn{1}{c}{\textbf{pre}} & \multicolumn{1}{c}{\textbf{rec}} & \multicolumn{1}{c}{\textbf{f1}} & \multicolumn{1}{c}{\textbf{fpr}} & \textbf{count} & \textbf{relevance}\\ \hline
\multirow{5}{*}{\textbf{Qwen3-8B}}   & \multicolumn{1}{c|}{\textbf{\textgreater{}0.95}} & 53.20& 53.97& 43.96& 48.45& 37.55   & 1423& \multirow{5}{*}{\begin{tabular}[c]{@{}c@{}}sc:-0.0192\\ {[}-0.0511,0.0126{]}\\ p:2.3079e-1\end{tabular}}  \\
 & \multicolumn{1}{c|}{\textbf{0.9-0.95}}& 51.81& 52.40& 41.16& 46.10& 37.50   & 689&\\
 & \multicolumn{1}{c|}{\textbf{0.85-0.9}}& 53.89& 55.50& 39.22& 45.96& 31.45   & 566&\\
 & \multicolumn{1}{c|}{\textbf{0.8-0.85}}& 54.81& 56.35& 41.52& 47.81& 31.98   & 343&\\
 & \multicolumn{1}{c|}{\textbf{0.75-0.8}}& 55.79& 57.29& 45.45& 50.69& 33.88   & 242&\\ \hline
\multirow{5}{*}{\textbf{Qwen2.5-Coder-7B-Instruct}}& \multicolumn{1}{c|}{\textbf{\textgreater{}0.95}} & 52.74& 53.23& 45.08& 48.82& 39.61   & 1424& \multirow{5}{*}{\begin{tabular}[c]{@{}c@{}}sc:-0.015\\ {[}-0.0462,0.0187{]}\\ p:3.4936e-1\end{tabular}}   \\
 & \multicolumn{1}{c|}{\textbf{0.9-0.95}}& 53.44& 54.12& 43.39& 48.17& 36.57   & 698&\\
 & \multicolumn{1}{c|}{\textbf{0.85-0.9}}& 52.84& 53.74& 40.78& 46.37& 35.11   & 564&\\
 & \multicolumn{1}{c|}{\textbf{0.8-0.85}}& 56.19& 58.68& 42.77& 49.48& 30.30   & 331&\\
 & \multicolumn{1}{c|}{\textbf{0.75-0.8}}& 52.42& 52.83& 45.16& 48.70& 40.32   & 248&\\ \hline
\multirow{5}{*}{\textbf{Llama-3.1-8B-Instruct}}  & \multicolumn{1}{c|}{\textbf{\textgreater{}0.95}} & 52.14& 52.18& 50.35& 51.25& 46.08   & 1427& \multirow{5}{*}{\begin{tabular}[c]{@{}c@{}}sc:-0.0249\\ {[}-0.0596,0.0009{]}\\ p:6.6968e-2\end{tabular}}  \\
 & \multicolumn{1}{c|}{\textbf{0.9-0.95}}& 52.19& 52.54& 45.19& 48.59& 40.82   & 686&\\
 & \multicolumn{1}{c|}{\textbf{0.85-0.9}}& 55.87& 56.37& 51.96& 54.07& 40.21   & 562&\\
 & \multicolumn{1}{c|}{\textbf{0.8-0.85}}& 54.60& 55.06& 50.00& 52.41& 40.80   & 348&\\
 & \multicolumn{1}{c|}{\textbf{0.75-0.8}}& 54.03& 54.39& 50.00& 52.10& 41.94   & 248&\\ \hline
\multirow{5}{*}{\textbf{deepseek-llm-7b-chat}}   & \multicolumn{1}{c|}{\textbf{\textgreater{}0.95}} & 51.50& 51.93& 40.38& 45.43& 37.38   & 1268& \multirow{5}{*}{\begin{tabular}[c]{@{}c@{}}sc:0.0027\\ {[}-0.0283,0.0352{]}\\ p:8.7198e-1\end{tabular}}   \\
 & \multicolumn{1}{c|}{\textbf{0.9-0.95}}& 51.36& 51.78& 39.46& 44.79& 36.75   & 664&\\
 & \multicolumn{1}{c|}{\textbf{0.85-0.9}}& 53.03& 54.55& 37.36& 44.35& 31.25   & 545&\\
 & \multicolumn{1}{c|}{\textbf{0.8-0.85}}& 51.65& 52.88& 32.93& 40.59& 29.52   & 333&\\
 & \multicolumn{1}{c|}{\textbf{0.75-0.8}}& 52.03& 52.58& 41.46& 46.36& 37.40   & 246&\\ \hline
\multirow{5}{*}{\textbf{deepseek-coder-6.7b-instruct}} & \multicolumn{1}{c|}{\textbf{\textgreater{}0.95}} & 50.68& 50.91& 50.15& 50.53& 48.79   & 1326& \multirow{5}{*}{\begin{tabular}[c]{@{}c@{}}sc:-0.0186\\ {[}-0.0519,0.0144{]}\\ p:2.6198e-1\end{tabular}}  \\
 & \multicolumn{1}{c|}{\textbf{0.9-0.95}}& 51.95& 52.49& 41.14& 46.13& 37.24   & 666&\\
 & \multicolumn{1}{c|}{\textbf{0.85-0.9}}& 50.10& 50.00& 43.30& 46.41& 43.13   & 523&\\
 & \multicolumn{1}{c|}{\textbf{0.8-0.85}}& 52.32& 53.54& 41.72& 46.90& 36.88   & 323&\\
 & \multicolumn{1}{c|}{\textbf{0.75-0.8}}& 53.14& 53.27& 47.90& 50.44& 41.67   & 239&\\ \hline
\end{tabular}
\label{table:finding2}
\end{table*}

In this RQ, we investigate how the semantic gap between a vulnerable code snippet and its corresponding patch influences the model's decision-making process.
As established in RQ1, models fine-tuned without explicit reasoning often fall into a "semantic trap," where they over-rely on superficial code patterns rather than capturing the intrinsic root causes of vulnerabilities. However, finetuning with explicit reasoning alleviates this issue, making the models noticeably more robust against such pitfalls.
Building on these findings, we hypothesize that the semantic gap (i.e., the degree of modification in a patch) is a decisive factor in this phenomenon: a smaller gap (high similarity) creates a stronger "trap" that misleads the model into treating both versions as identical, whereas a larger gap provides more distinct structural signals for the model to differentiate between vulnerable and fixed states.

To test this hypothesis, we reorganize \DA into distinct training and test splits where each vulnerable sample is strictly paired with its corresponding patch in the same split. This ensures that evaluation always occurs on complete pairs (vuln, patch).
We then fine-tune the five representative LLMs on this restructured \DA under two paradigms. We evaluate their performance across subsets of test pairs stratified by CodeBLEU score, a metric introduced in \S~\ref{sec:codebleu} to quantify semantic similarity between code fragments.
The test samples are divided into five distinct intervals based on their CodeBLEU scores: (1) \textgreater0.95, (2) 0.90–0.95, (3) 0.85–0.90, (4) 0.80–0.85, and (5) 0.75–0.80, with lower scores indicating greater semantic divergence. Pairs with a CodeBLEU score \textless 0.75 are excluded, as there are fewer than 325 pairs in that range, which is not statistically significant when split into intervals.

Furthermore, to rigorously measure the statistical dependency between the semantic gap (CodeBLEU score) and the models' detection metrics, we calculated the Spearman's rank correlation coefficient (\textit{sc}). We also report the 95\% confidence intervals and p-values to evaluate the statistical significance of this correlation across both fine-tuning paradigms. The comprehensive performance trends and statistical analyses are presented in Table~\ref{table:finding2}.

As shown in Table~\ref{table:finding2}, the relationship between the semantic gap (measured by CodeBLEU) and model performance differs depending on the fine-tuning paradigm.
Under the without-explicit-reasoning paradigm, performance metrics exhibit a consistent upward trend as the semantic divergence between vulnerable and patched code increases (i.e., as CodeBLEU scores decrease). For instance, \texttt{Qwen3-8B}'s F1-score improves from 50.41\% (CodeBLEU \textgreater0.95) to 63.07\% (CodeBLEU 0.75–0.80), an increase of over 12 percentage points. Similarly, \texttt{deepseek-coder-6.7b-instruct} sees its F1 rise from 41.91\% to 58.77\%.
This stratification-level trend is supported by our statistical analysis: for all five models under the without-explicit-reasoning paradigm, the Spearman's rank correlation coefficients (\textit{sc}) are strictly negative, their 95\% confidence intervals lie entirely within the negative range, and all p-values are well below the 0.05 significance threshold (e.g., $p = 3.4084e^{-6}$ for \texttt{Qwen3-8B} and $p=3.3016e^{-5}$ for \texttt{deepseek-coder-6.7b-instruct}).
We emphasize, however, that the \emph{magnitude} of this correlation is small ($|\textit{sc}|$ ranges from $0.033$ to $0.074$, all below the conventional $|\rho|=0.1$ threshold for a weak effect). The large sample size per stratum is what drives statistical significance, not a strong dependency at the sample level. Accordingly, we interpret the result as a \textit{weak but systematic} bias: the semantic gap is one but not the dominant factor shaping a non-reasoning model's prediction on (vulnerable, patched) pairs. We also acknowledge an inherent confound: pairs with a higher CodeBLEU score differ by fewer security-critical tokens and are therefore intrinsically harder to discriminate for any classifier, including a human expert; the symptom we attribute to the \textit{Semantic Trap} (symptom~(ii) in §I) is the model's \emph{reliance} on this gap as a discriminative cue rather than the existence of difficulty alone.

In contrast, models fine-tuned under the with-explicit-reasoning paradigm do not exhibit a statistically detectable dependency on CodeBLEU. The F1-scores fluctuate non-monotonically across CodeBLEU intervals; for example, the F1-score of \texttt{Qwen2.5-Coder-7B-Instruct} moves from 48.82\% (\textgreater0.95) to 46.37\% (0.85–0.90) and back to 49.48\% (0.80–0.85).
The \textit{sc} values are near zero, the 95\% confidence intervals all cross zero, and the p-values consistently exceed 0.05 (e.g., $p=0.349$ for \texttt{Qwen2.5-Coder-7B-Instruct} and $p=0.871$ for \texttt{deepseek-llm-7b-chat}), indicating that under this paradigm we cannot reject the null hypothesis of no correlation between the semantic gap and the model's detection metrics.

We are careful, however, not to over-interpret this absence of correlation as direct evidence that explicit reasoning ``escapes'' the Semantic Trap. Two alternative explanations remain compatible with the data and should be ruled out before drawing such a causal conclusion. First, a Fisher $z$-test comparing the two paradigms' correlations on a per-model basis reveals that the gap between $\textit{sc}_{\text{w.o.}}$ and $\textit{sc}_{\text{w.}}$ is itself statistically significant only for a subset of models, leaving open the possibility that the paradigm-level contrast is partly driven by sampling variation. Second, and more importantly, the with-explicit-reasoning models exhibit substantially lower recall (Table~\ref{table:finding1}), pushing their F1 toward the $\sim$50\% region across all CodeBLEU strata; such a compressed dynamic range can attenuate any underlying correlation through a \emph{floor effect}, mimicking the appearance of paradigm-induced decoupling without the model actually reasoning over the security logic. The reasoning-failure analysis in RQ3 corroborates that with-explicit-reasoning models still systematically misinterpret control flow and API behavior, suggesting that any mitigation of the trap is partial at best. Taken together, our results support a more measured claim: under the without-explicit-reasoning paradigm we observe a weak but statistically systematic gap-dictated bias consistent with symptom~(ii) of the Semantic Trap, while the with-explicit-reasoning paradigm reduces but does not demonstrably eliminate this dependency.

\finding{}{
Under the without-explicit-reasoning paradigm, a fine-tuned model's detection performance on (vulnerable, patched) pairs is weakly but systematically biased by the superficial semantic gap (CodeBLEU): Spearman $|\textit{sc}|<0.1$ across all five models with $p<0.05$, consistent with symptom~(ii) of the Semantic Trap. The with-explicit-reasoning paradigm reduces the measurable dependency to a statistically non-significant level, but the result must be interpreted cautiously: the absence of correlation is also compatible with a floor effect induced by the lower recall of these models, and reasoning failures persist (RQ3), suggesting that explicit reasoning mitigates rather than removes the trap.
}

\subsection{Demystifying the Semantic Trap: An In-depth Analysis of Reasoning Failures (RQ3)}

\begin{table*}[]
\caption{Taxonomy and detailed definitions of the eight distinct reasoning failure categories identified in LLMs during vulnerability detection.}
\begin{tabularx}{\linewidth}{|c|>{\centering\arraybackslash}p{0.25\linewidth}|>{\centering\arraybackslash}X|}
\hline
\textbf{ID} & \textbf{name} & \textbf{Definition} \\ \hline
\textbf{C1}& \textbf{Control Flow and Execution Path Misinterpretation}    & The model fails to correctly trace the logical sequence of code execution, leading to incorrect conclusions about whether specific lines of code are reachable, the order of operations, or the impact of conditional branches (e.g., early returns, short-circuit evaluation).    \\ \hline
\textbf{C2}& \textbf{Library, API, and Framework Semantics Hallucination}  & The model demonstrates a fundamental lack of knowledge regarding the specific behavior, safety guarantees, or return values of standard library functions, kernel helpers, or framework APIs, often assuming they are unsafe or behave differently than they actually do.\\ \hline
\textbf{C3}& \textbf{Contextual and Scope Isolation Errors}      & The model analyzes a code snippet in isolation, failing to account for broader architectural context, caller guarantees, upstream/downstream validation, or the specific environment (e.g., kernel vs. userspace, test harness vs. production) in which the code operates. \\ \hline
\textbf{C4}& \textbf{Data Type and Arithmetic Logic Flaws}& The model exhibits misunderstandings of programming language semantics regarding data types (signed vs. unsigned, fixed-width integers), arithmetic operations (overflow, underflow, wraparound), and memory layout (padding, struct alignment).   \\ \hline
\textbf{C5}& \textbf{Vulnerability Category and Definition Misapplication} & The model correctly identifies a potential issue but maps it to the wrong Common Weakness Enumeration (CWE) category, or identifies a theoretical risk that does not meet the specific criteria of the requested vulnerability type (e.g., confusing Input Validation with Buffer Overflow). \\ \hline
\textbf{C6}& \textbf{Concurrency and Lifecycle Management Blind Spots}     & The model fails to accurately assess race conditions, reference counting mechanisms, object lifecycles, and synchronization primitives, often hallucinating race conditions where locks exist or missing Use-After-Free (UAF) scenarios due to complex ownership transfers.\\ \hline
\textbf{C7}& \textbf{Input Trust Boundary and Sanitization Misjudgment}    & The model incorrectly determines the trustworthiness of data sources, either treating sanitized/internal data as malicious or failing to recognize that certain inputs (e.g., hardware registers, network packets) require strict validation that is absent.  \\ \hline
\textbf{C8}& \textbf{Speculative and Hallucinated Code Analysis} & The model invents code behaviors, variables, function calls, or execution contexts that do not exist in the provided snippet, basing its reasoning on fabricated premises rather than the actual source code.\\ \hline
\end{tabularx}
\label{table:rq3}
\end{table*}

\begin{figure*}[htbp]
\centering
\includegraphics[width=\textwidth]{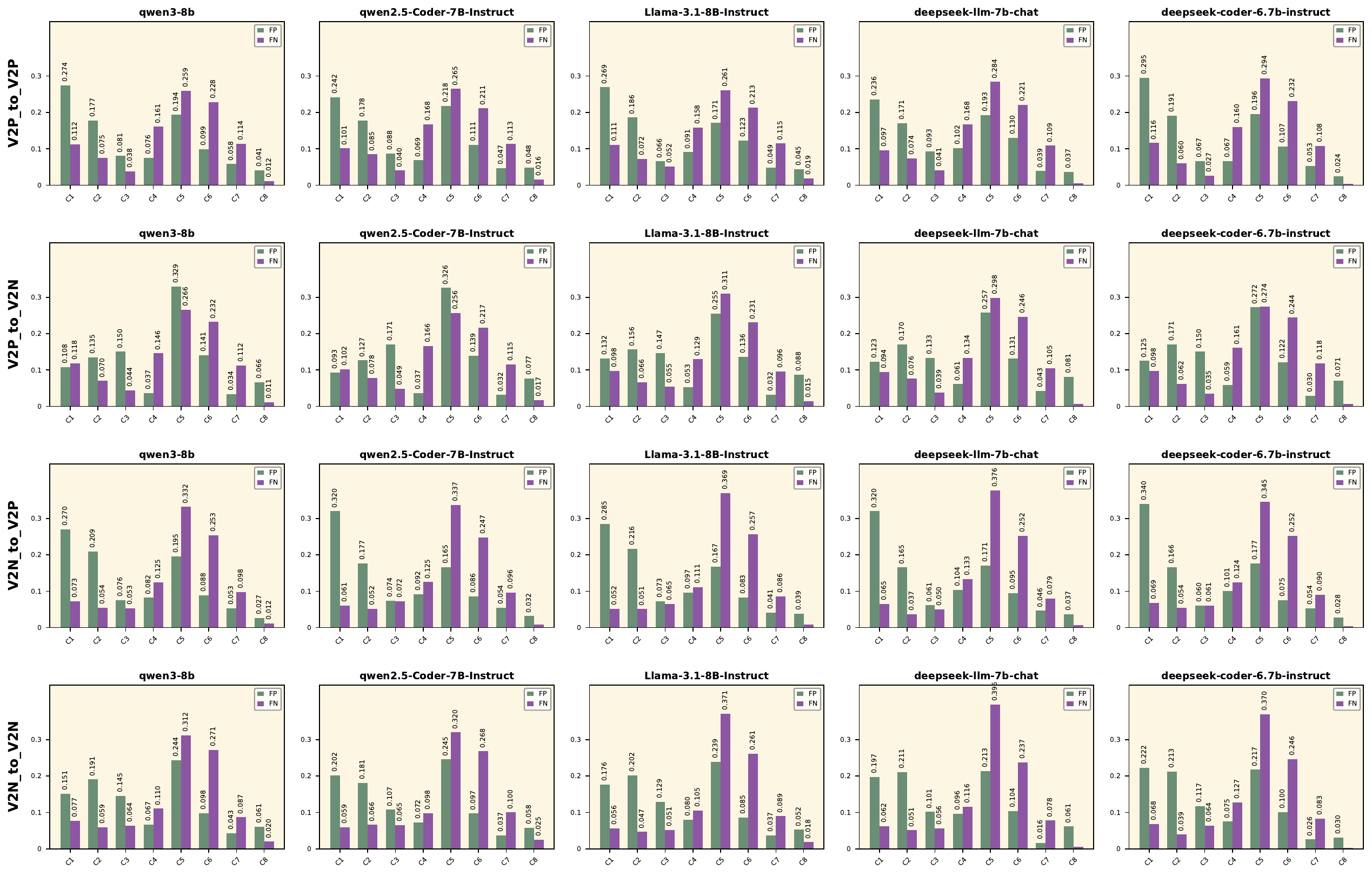}
\caption{Distribution of reasoning failure causes across different models and experimental configurations. In each subplot, the x-axis indicates the error category IDs (C1–C8) as defined in Table~\ref{table:rq3}, and the y-axis represents the proportion of each error category within the specific experimental setup.}
\label{fig:rq3}
\end{figure*}

In this RQ, we delve deeper into the behavior of models fine-tuned under the explicit-reasoning paradigm. RQ1 and RQ2 showed that this paradigm reduces the measurable symptoms of the Semantic Trap but does not eliminate failed predictions. To understand the kinds of failures that remain, we analyze the explicit-reasoning traces produced alongside each incorrect prediction (i.e., False Positives and False Negatives). We emphasize an important methodological point that determines what RQ3 can and cannot show: because the fine-tuned model's final label is binary (\textit{vulnerable}/\textit{safe}), categories such as ``CWE misapplication'' (C5) or ``concurrency blind spot'' (C6) are not directly observable from the label alone, they are derived by an evaluator LLM that reads the natural-language reasoning text the model emits under the explicit-reasoning prompt template (Figure~\ref{fig:promptTemplate}) and compares it against the ground-truth reasoning distilled in \S\ref{sec:dataset}. Consequently, RQ3 is best understood as characterizing the distribution of \emph{failure descriptions} produced by this evaluator pipeline, not as a direct measurement of the fine-tuned model's internal cognition.

Our analytical approach consists of two main steps. First, to construct a systematic taxonomy of reasoning errors, we employ an advanced LLM to act as an expert code security evaluator. For each failed sample, we provide the evaluator LLM with a comprehensive context: the original code snippet, the ground-truth reasoning, and the fine-tuned model's flawed reasoning output. We prompt the LLM to comparatively analyze these inputs and pinpoint the specific logical flaws, misinterpretations, or missing considerations that led to the incorrect conclusion. From this comprehensive pool of failure explanations, we randomly sample 3,000 texts. We then prompt the LLM to synthesize and summarize these samples, deriving a structured categorization of reasoning failures. This process yielded eight distinct reasoning error categories (C1–C8), with detailed definitions presented in Table~\ref{table:rq3}.

Subsequently, to quantify the prevalence of these reasoning blind spots, we utilize the LLM once more to evaluate all failed samples across our experiments, mapping each individual failure to its most appropriate category in our established taxonomy. Finally, we analyze and compare the distribution of these reasoning error causes across various experimental configurations and different fine-tuned models. The resulting statistical distributions, highlighting both False Positives (FP) and False Negatives (FN) for each category, are visualized in Figure~\ref{fig:rq3}.

As detailed in Table~\ref{table:rq3}, we leveraged an advanced LLM to systematically categorize the root causes of incorrect reasoning generated by the fine-tuned models into eight distinct types. These encompass Control Flow and Execution Path Misinterpretation (C1), Library/API and Framework Semantics Hallucination (C2), Contextual and Scope Isolation Errors (C3), Data Type and Arithmetic Logic Flaws (C4), Vulnerability Category and Definition Misapplication (C5), Concurrency and Lifecycle Management Blind Spots (C6), Input Trust Boundary and Sanitization Misjudgment (C7), and Speculative and Hallucinated Code Analysis (C8).

Subsequent analysis of the error distributions presented in Figure~\ref{fig:rq3} yields several critical observations regarding how and why these models fail. Primarily, the distribution patterns of reasoning error causes are remarkably similar across different models and experimental configurations (e.g., \texttt{V2P\_to\_V2P} versus \texttt{V2N\_to\_V2N}). This consistency indicates that these reasoning flaws are systemic limitations inherent to the LLMs' current capabilities rather than artifacts of specific training data compositions.

Despite this overarching consistency, there is a significant divergence in the primary causes driving False Positives (FPs) compared to False Negatives (FNs). For FPs, the proportions of \textbf{C1} and \textbf{C2} errors are substantially higher than those observed in FNs, demonstrating that control flow misinterpretation and API hallucination are the leading causes of false alarms. For example, under the \texttt{V2N\_to\_V2P} setting for \texttt{qwen2.5-Coder-7B-Instruct}, C1 and C2 account for 32.0\% and 17.7\% of FP errors, respectively, compared to a mere 6.1\% and 5.2\% in FN errors. Conversely, when the models fail to detect actual vulnerabilities (FNs), the proportions of \textbf{C5} and \textbf{C6} generally dominate. Taking \texttt{deepseek-coder-6.7b-instruct} in the \texttt{V2N\_to\_V2N} setting as an illustration, C5 and C6 constitute a massive 37.0\% and 24.6\% of FN errors, far outweighing their contributions to FPs (21.7\% and 10.0\%). This highlights that vulnerability category misapplication and concurrency blind spots are the primary reasons models miss real threats. Furthermore, comprehensively observing all scenarios in Figure 3 reveals that \textbf{C5} consistently exhibits the highest frequencies across various conditions, establishing it as the most universally prevalent cause of reasoning failure overall.

Synthesizing these empirical observations, we find that the most frequent failure categories are consistent with shallow understanding of specific programming environments (such as C/C++ and the Linux kernel) rather than deep reasoning about security-critical logic. The high C1 and C2 rates in FPs are consistent with hallucinated API behavior and misread control flow, and the patterns in the failure texts often look like heuristic pattern matching (e.g., flagging the absence of an explicit check without verifying whether that check is implicit, delegated, or logically unnecessary). The dominance of C5 in FNs further indicates that models frequently confuse general robustness issues or standard logical bugs with security-critical memory-safety flaws, suggesting an imprecise grasp of CWE definitions and boundaries. We emphasize that these observations describe the distribution of error \emph{categories} produced by a single LLM-as-judge pipeline and should not be read as a direct measurement of the underlying cognitive mechanism; the next section discusses the corresponding threats to validity.

\finding{}{
False Positives are dominated by errors consistent with control-flow misinterpretation and API hallucination (C1, C2), whereas False Negatives are dominated by errors consistent with vulnerability-category misapplication (C5) and concurrency blind spots (C6). C5 is the most frequent category overall. These distributions are consistent with shallow semantic understanding and reliance on heuristic patterns, although the categorization is produced by an LLM evaluator and should be interpreted as a description of observable failure patterns rather than as a direct probe of the model's internal reasoning.
}

%% file: Sections/5-Discussion.tex
\section{Discussion}

\subsection{Threats to validity}
\label{sec:threats}

Threats to validity in our study primarily stem from the size of the evaluated models and the potential noise inherent in the vulnerability datasets.
Regarding the size of the models, our evaluation focuses on open-source LLMs within the 7B–8B parameter range due to computational resource constraints. We acknowledge that larger-scale models might demonstrate enhanced general reasoning capabilities. However, as corroborated by recent study~\cite{weissberg2025llm}, the fundamental bottleneck in vulnerability detection lies in the learning paradigm rather than mere model capacity. The ``Semantic Trap" illustrates a systemic flaw where models rely on superficial functional patterns instead of causal security logic. Simply scaling up the parameter count is unlikely to resolve this intrinsic limitation without significant shifts in model architectures or semantic-aware training methodologies.
Regarding potential dataset noise, it is well-recognized that commit-based datasets like PrimeVul can contain inaccurate labels. To mitigate this, we implemented a data cleaning pipeline, including hash-based deduplication and a priority-based label resolution strategy across multiple high-quality sources. This mitigates the impact of inherent noise on the model's training and evaluation.

\subsection{Lessons Learned}

Our investigation into the ``Semantic Trap" provides critical insights for the future of LLM-based vulnerability detection. We summarize the key lessons as follows:

\textbf{Fine-tuning is not a Panacea for Logic Comprehension.} A central observation of this study is that improving performance metrics through fine-tuning does not necessarily equate to a deeper understanding of vulnerability root causes. As demonstrated in RQ1, fine-tuned models can achieve deceptively high accuracy on traditional V2N settings but exhibit substantial robustness degradation under minor semantic-preserving data augmentations. Furthermore, our failure analysis in RQ3 shows that even when models are forced to generate step-by-step reasoning, they frequently fall back on patterns consistent with heuristic matching, hallucinating API behaviors (C2) or misapplying vulnerability definitions (C5). This suggests that the current fine-tuning paradigm inadvertently trains models to behave as ``sophisticated pattern matchers'' rather than ``logical reasoners.'' Therefore, researchers should be cautious of high accuracy scores on standard benchmarks, as they may mask limited comprehension of security-critical logic.

\textbf{The Criticality of V2P vs.\ V2N Cross-Evaluation.}
Our empirical results highlight a clear contrast between evaluating models on unpaired code (V2N) versus paired code (V2P). As shown in RQ1, models that perform well in V2N-to-V2N scenarios suffer a large performance drop when faced with V2P pairs, often producing high false-alarm rates on patched code. RQ2 further quantifies this phenomenon: under standard fine-tuning, a model's detection performance on (vulnerable, patched) pairs shows a weak but statistically systematic dependency on the superficial semantic gap (CodeBLEU) between the two snippets, consistent with the model exploiting text-modification magnitude as a partial discriminative signal. However, most existing research evaluates models only on unpaired datasets, leaving potential risks in real-world deployment unchecked. We therefore advocate for a routine cross-evaluation protocol that includes both V2P and V2N datasets when assessing whether a model has grasped fine-grained security logic or is partially relying on functional patterns.

\textbf{Moving Towards Code-Semantic-Aware Fine-Tuning.}
To escape the "Semantic Trap," future work must bridge the gap between statistical token learning and structured program logic. We suggest that fine-tuning should not rely on code alone. Instead, integrating explicit structural information, such as data flow and control flow information, or security-related execution paths, is essential to guide LLMs toward learning the "causal logic" of vulnerabilities. By forcing LLMs to attend to the relationships between variables and execution states rather than just keywords, LLMs will be able to learn how the code looks to how the code behaves from a security perspective.

%% file: Sections/7-Conclusion.tex
\section{Conclusion}

To summarize, in this paper we investigate whether fine-tuned LLMs genuinely learn vulnerability root causes or instead rely on functional patterns in vulnerability detection, and we identify a failure mode that we term the \textit{Semantic Trap}, characterized by three observable symptoms (pairing-sensitive performance, gap-dictated decisions, and fragility to semantic-preserving change). Using TrapEval, an evaluation framework that combines paired (V2P) and unpaired (V2N) datasets, semantic-preserving perturbations, semantic-gap analysis, and an LLM-assisted error taxonomy, we conduct an extensive study across five state-of-the-art LLMs under two fine-tuning paradigms (with and without explicit reasoning).
Our findings highlight three key observations: (1) models trained under the without-explicit-reasoning paradigm achieve deceptively high performance on traditional datasets (V2N) but suffer large false-alarm rates on patched code (V2P) and noticeable robustness degradation under semantic-preserving transformations; (2) the detection performance of these models on (vulnerable, patched) pairs shows a weak but statistically systematic dependency on the superficial semantic gap (CodeBLEU), and although fine-tuning with explicit reasoning reduces this measurable dependency, the result is partly compatible with a floor effect arising from lower recall and should not be read as direct evidence of escaping the trap; and (3) explicit-reasoning models still frequently fail in ways consistent with shallow semantic understanding, predominantly misinterpreting control flow (driving False Positives) and misapplying vulnerability definitions (driving False Negatives).
Taken together, these results indicate that current fine-tuning practices tend to reinforce shortcut learning rather than imparting genuine security reasoning. We conclude that paired evaluations are essential for accurately assessing LLM-based vulnerability detectors, and that future research should move toward code-semantic-aware training paradigms to enable reliable automated software security analysis.

\section{Acknowledgement}
This work was supported in part by National Key R\&D Program of China (2023YFB3106800), by the National Natural Science Foundation of China (U25B2024, 62572437), by the National Natural Science Foundation of Hangzhou (No.2025SZRJJ1712).